\documentclass{aa}  
\usepackage{graphicx}
\usepackage{amsmath}
\usepackage{txfonts}
\begin{document}

 \title{Using Parker Solar Probe observations during the first four perihelia to constrain global magnetohydrodynamic models}
 
 \titlerunning{Comparing PSP observations with MHD models}
 \authorrunning{Riley et al.}


   \author{Pete Riley\inst{1}, Roberto Lionello\inst{1}, Ronald M. Caplan\inst{1}, Cooper Downs\inst{1},  Jon A. Linker\inst{1}, Samuel T. Badman\inst{2,3}
          \and 
   Michael L. Stevens\inst{4} }

   \institute{Predictive Science Inc., San Diego, California, USA \email{pete@predsci.com}
 	 \and
  	Physics Department, University of California, Berkeley, CA 94720-7300, USA        
   	\and
   	Space Sciences Laboratory, University of California, Berkeley, CA 94720-7450, USA
   	\and
   	Smithsonian Astrophysical Observatory, Cambridge, MA 02138, USA
   }    

   \date{Received 30 October 2020; accepted N/A}

 
  \abstract
   {Parker Solar Probe (PSP) is providing an unprecedented view of the Sun's corona as it progressively dips closer into the solar atmosphere with each solar encounter. Each set of observations provides a unique opportunity to test and constrain global models of the solar corona and inner heliosphere and, in turn, use the model results to provide a global context for interpreting such observations.}
   {In this study, we develop a set of global magnetohydrodynamic (MHD) model solutions of varying degrees of sophistication for PSP's first four encounters and compare the results with in situ measurements from PSP, Stereo-A, and Earth-based spacecraft, with the objective of assessing which models perform better or worse. We also seek to understand whether the so-called `open flux problem', which all global models suffer from, resolves itself at closer distances to the Sun.}
   {The global structure of the corona and inner heliosphere is calculated using three different MHD models. The first model (``polytropic''), replaced the energy equation as a simple polytropic relationship to compute coronal solutions and relied on an ad hoc method for estimating the boundary conditions necessary to drive the heliospheric model. The second model (``thermodynamic'') included a more sophisticated treatment of the energy equation to derive the coronal solution, yet it also relied on a semi-empirical approach to specify the boundary conditions of the heliospheric model. The third model (``WTD'') further refines the transport of energy through the corona, by implementing the so-called wave-turbulence-driven approximation. With this model, the heliospheric model was run directly with output from the coronal solutions. All models were primarily driven by the observed photospheric magnetic field using data from Solar Dynamics Observatory's  Helioseismic and Magnetic Imager (HMI) instrument. }
   {Overall, we find that there are substantial differences between the model results, both in terms of the large-scale structure of the inner heliosphere during these time periods, as well as in the inferred timeseries at various spacecraft. The ``thermodynamic'' model, which represents the ``middle ground'', in terms of model complexity, appears to reproduce the observations most closely for all four encounters. Our results also contradict an earlier study that had hinted that the open flux problem may disappear nearer the Sun. Instead, our results suggest that this ``missing'' solar flux is still missing even at $26.9 R_S$, and thus it cannot be explained by interplanetary processes. Finally, the model results were also used to provide a global context for interpreting the localized in situ measurements.}
   {Earlier studies suggested that the more empirically-based polytropic solutions provided the best matches with observations. The results presented here, however, suggest that the thermodynamic approach is now superior. We discuss possible reasons for why this may be the case, but, ultimately, more thorough comparisons and analyses are required. Nevertheless, it is reassuring that a more sophisticated model appears to be able to reproduce observations since it provides a more fundamental glimpse into the physical processes driving the structure we observe.}


\maketitle
%

\section{Introduction}

  NASA's Parker Solar Probe (PSP) spacecraft launched on 12 August 2018 and reached its first of 24 perihelia (P1) on 5 November 2018. Since then it has successfully completed six perihelion encounters (as of 27 September 2020), with ever decreasing distances of closest approach. PSP's primary scientific goals are to:  (1) better understand what heats the solar corona and accelerates the solar wind; (2) determine the underlying structure and dynamics of the coronal magnetic field; and (3) better identify the mechanisms that accelerate and transport energetic particles in the corona \citep{fox16a}. 
  
PSP carries four instrument packages. Of these, two are particularly relevant for studying the large-scale magnetic and plasma properties of the solar corona and inner heliosphere. FIELDS (Electro- magnetic Fields Investigation) consists of two flux-gate magnetometers, a search-coil magnetometer, and five plasma voltage sensors \citep{bales16a}. It measures electric and magnetic fields, as well as radio waves, Poynting flux, plasma density, and electron temperature. Solar Wind Electrons Alphas and Protons (SWEAP) is composed of three instruments: two electrostatic analyzers and one Faraday cup, from which estimates of velocity, density, and temperature of electrons, protons, and alpha particles can be made \citep{kasper16a}.
  
Global models of the solar corona and inner heliosphere can provide crucial support for interplanetary missions \citep[e.g.,][]{riley01a,torok18a}. Not only do they provide a global picture of the properties and structure of the heliosphere, but they allow observers to connect the otherwise disparate observations to one another. In addition to computing basic magnetic and plasma variables, models can also be used to reconstruct convolved measurements, such as extreme ultraviolet (EUV) and white light images \citep{mikic18a}. 

A number of numerical models have been developed and applied to help in the interpretation of PSP observations. Potential Field Source Surface (PFSS) models represent the simplest approach, where: (1) the corona is assumed to be current free; (2) time-dependent effects are ignored (or treated as a series of quasi-equilibria); and (3) a source surface outer boundary condition is imposed, where the field lines are forced to become radial (typically set to be $2.5 R_S$ \citep[e.g.,][]{badman20a}). In spite of their inherent simplicity, PFSS models have been shown to compare well with MHD solutions, at least under some conditions \citep{riley06b}. \citet{badman20a} demonstrated the utility of applying PFSS models to interpret the magnetic connectivity between PSP and the solar surface, and, in particular, in being able to predict the location of these foot-points. Of course, the PFSS approach is limited to inquiries concerning the magnetic structure, and cannot directly address the plasma properties within the corona or further out. 

\citet{holst19a} applied the Alfv\'en Wave Solar atmosphere Model (AWSoM) to predict the in situ measurements that would be returned from PSP during P1. In this model, outwardly propagating low-frequency Alfv\'en waves, which are partially reflected, provide both the coronal heating and acceleration. Although no comparisons were made with PSP in that study, a visual comparison with OMNI data at 1 AU, suggests that the model had captured the large-scale stream structure of the solar wind. A qualitative comparison with the predictions and subsequent observations at PSP supports the general state of the solar wind (slow, dense) at perihelion, although the modest structure that was apparent was not reproduced by the model. Moreover, the model underestimated the strength of the magnetic field. In this case, they relied on ADAPT-GONG maps, which include a more modest 1.85 correction factor to amplify the photospheric magnetic fields. This need to boost either the photospheric magnetic field boundary conditions, or the resulting model fields at 1 AU, is a consistent requirement for all global MHD models, and, depending on the input magnetogram, can range from 1.5 to 3.0. 

We also developed a 3D wave-turbulence-driven (WTD) MHD prediction for the state of the corona and inner heliosphere for P1. The model was driven by photospheric magnetic field observations several weeks prior to the encounter \citep{riley19a}, thus, it was a true prediction. This served as both as a test of our model's predictive capabilities as well as what we hoped would be an aid in mission planning. For example, it would inform ground-based observers where they should target their smaller-scale campaign observations, based on our predictions of the foot-point locations during perihelion. We inferred that, in the days prior to first encounter, PSP would be immersed in wind emanating from a well-established, positive-polarity northern polar coronal hole. During the encounter, however, field lines from the spacecraft would map to a negative-polarity equatorial coronal hole, within which it would remain for the entire encounter, before becoming magnetically connected to a positive-polarity equatorial coronal hole. In that study, we also compared MHD and PFSS predictions, noting that while there are an overall agreement in the forecasts, there were some notable differences. The equatorial coronal holes, for example, predicted by the PFSS solutions were substantially smaller than those inferred from the MHD model results. 

Most recently, \citet{reville20a} modified the astrophysical Pluto magnetohydrodynamic  (MHD) code \citep{mignone07a} by accounting for Alfv\'en wave transport and dissipation. The wave-turbulence approach was similar to the previously described studies \citep{riley19a,holst19a} but simpler in that the Alfv\'en reflection process was not explicitly included. Nevertheless, they found that, at least for P1, the large-scale amplitude of the plasma parameters were well reproduced by the model. Although the modeled values of the magnetic field strength could be interpreted as being an underestimate, they argued that, in fact, if the amplitude of the waves used to heat the corona and accelerate the solar wind are added post hoc to the radial and total magnetic fields, this would bring the model results into agreement with the observations. While intriguing, this requires more substantiation through comparisons with subsequent perihelia, at different spacecraft locations, and with simultaneous comparison with other observational metrics, such as EUV and white-light observations, 

Finally, bridging the gap between the fully MHD and PFSS approaches, \citet{kim20a} presented model results for each of the first three PSP perihelia using a model composed of a coronal PFSS model, connected to a heliospheric MHD model. They used the Wang-Sheelely-Arge (WSA) methodology for prescribing the solar wind speed at the inner boundary of the heliospheric model \citep{arge03a}, which gives largely similar results to the Distance from the Coronal Hole (DCHB) method that we have developed \citep{riley01a,riley15a}. The most significant disagreements arise in the vicinity of pseudo-streamers, where the WSA model predicts slow solar wind, in contradiction to observations.  

In this study, we compare MHD results from three distinct approaches with in situ measurements made by PSP, Stereo-A, and Earth-based spacecraft (ACE and Wind) in an effort to identify any systematic differences between the model results. Based on this, we then use the best model results to infer the global structure of the heliosphere during each of the first four perihelia encounters. Additionally, we highlight how sensitive the timeseries comparisons are to the precise trajectory of the spacecraft through the model solution, suggesting that even when differences exist, the overall global structure may still be reasonably accurate. Finally, given PSP's ever smaller point of closest approach we interpret the comparisons in terms of whether an ``open flux problem'' remains. 

 \begin{figure*}
\centering
\includegraphics[width=18cm]{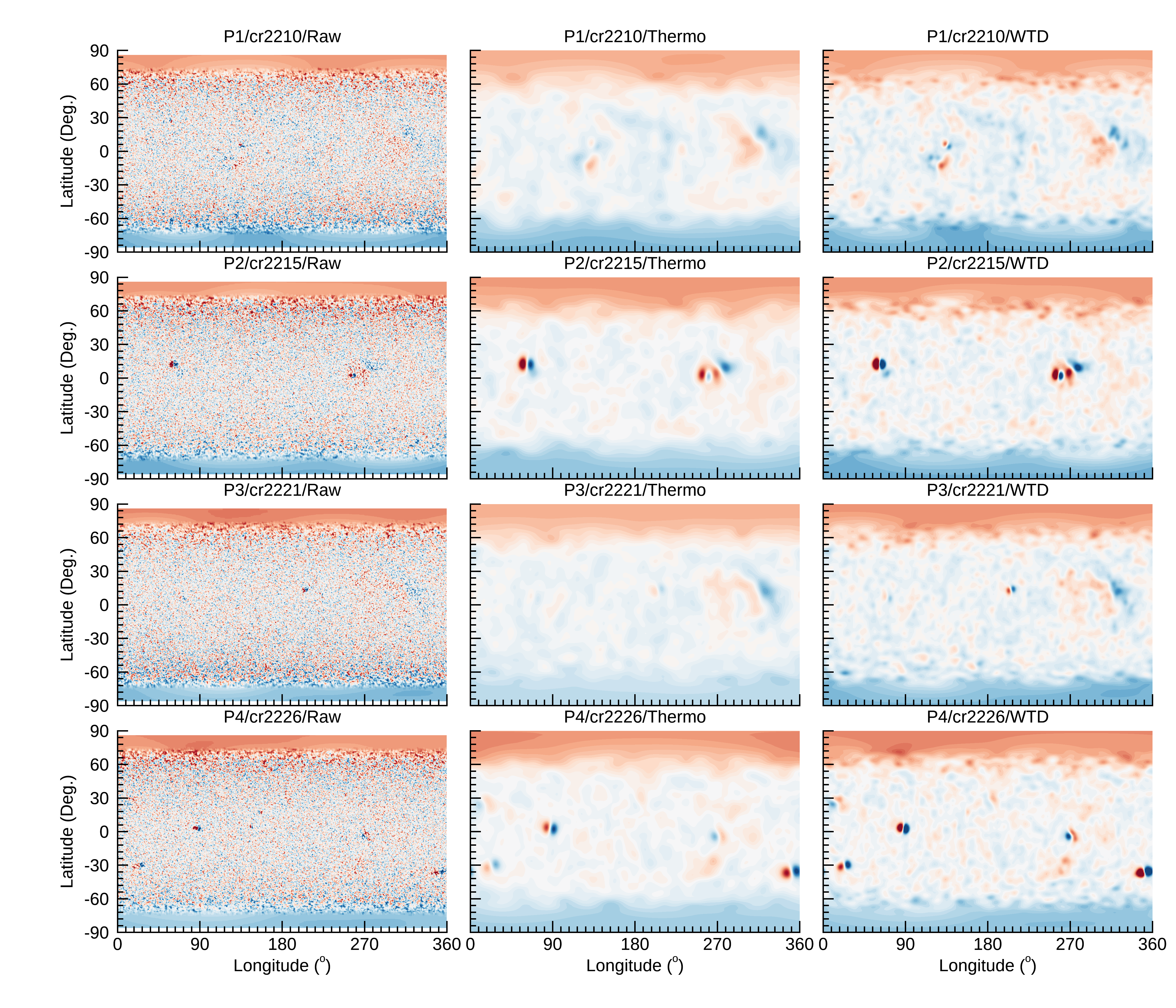}
\caption{Comparison of HMI magnetograms for each for the four Perihelia intervals (P1, P2, P3, and P4), which occurred during CR 2210, 2215, 2221, and 2226. In the left column are the original HMI synoptic maps, with the polar regions filled in.  In the middle column are the processed magnetograms used to drive the polytropic and thermodynamic models, and in the right column are the synoptic maps used to drive the WTD model. Blue/red corresponds to negative/positive polarities, respectively, and the maps have been saturated at $\pm 10$ G; thus, some of the ARs such as during CR2215, contain field strengths larger than this maximum. 
}
\label{f:magnetograms}
\end{figure*}

\section{Methods}

\subsection{Data}

Data used to drive the model were obtained from the Helioseismic and Magnetic Imager (HMI) onboard the Solar Dynamics Observatory (SDO) spacecraft \citep{scherrer12a}, and, specifically, from jsoc.stanford.edu/ajax/exportdata.html. These data provide an estimate for the radial component of the field on a uniform grid size of $3600 \times 1440$ in longitude and sin(latitude), respectively. 

Figure~\ref{f:magnetograms} summarizes each synoptic map for CR 2200, 2215, 2221, and 2226, which contained each of the perihelia P1, P2, P3, and P4, respectively. We note several points. First, to produce MHD solutions, the raw magnetograms (left-most column) must be processed to: (1) remove small-scale features (e.g., parasitic polarities) that are below the resolution of the model; and (2) provide a reasonable estimate for the polar fields, which are poorly, if at all observed. As can be seen from the middle and right-most columns, the two processing pipelines we currently use generate similar but not identical processed maps. The middle column relies on our simpler, but mature technique that is used to develop our online, standard solutions (e.g., \url{http://www.predsci.com/mhdweb/home.php}). Flux is balanced and preserved, meridional variations are made smoother, but the underlying structure visible in the original map remains. The right-most column relies on a more recent approach to developing input magnetograms for the MAS code. It attempts to maintain more structure , particularly at higher latitudes. Perhaps most importantly, for the maps for the polytropic and thermodynamic runs, the poles are filled in by extrapolation of the mid-latitude fields, while for the WTD runs we use the pole-filled data provided by the HMI instrument team. For our purposes, however, the main point is that they serve as different, but potentially equally-valid drivers of the MAS code. 
   
%

In this study, we use in situ measurements from ACE and Wind, in the form of the merged OMNI dataset as well as from Stereo-A, and, of course PSP. From each spacecraft we compare solar wind bulk speed, proton number density, and radial magnetic field. Together, these three quantities characterize the basic dynamical properties of the solar wind. All data were obtained from NASA's Space Physics Data Facility (SPDF) web services API \citep[e.g.,][]{candey19a}, and were retrieved at 1-hour resolution.

 \begin{figure*}
\centering
\includegraphics[width=17cm]{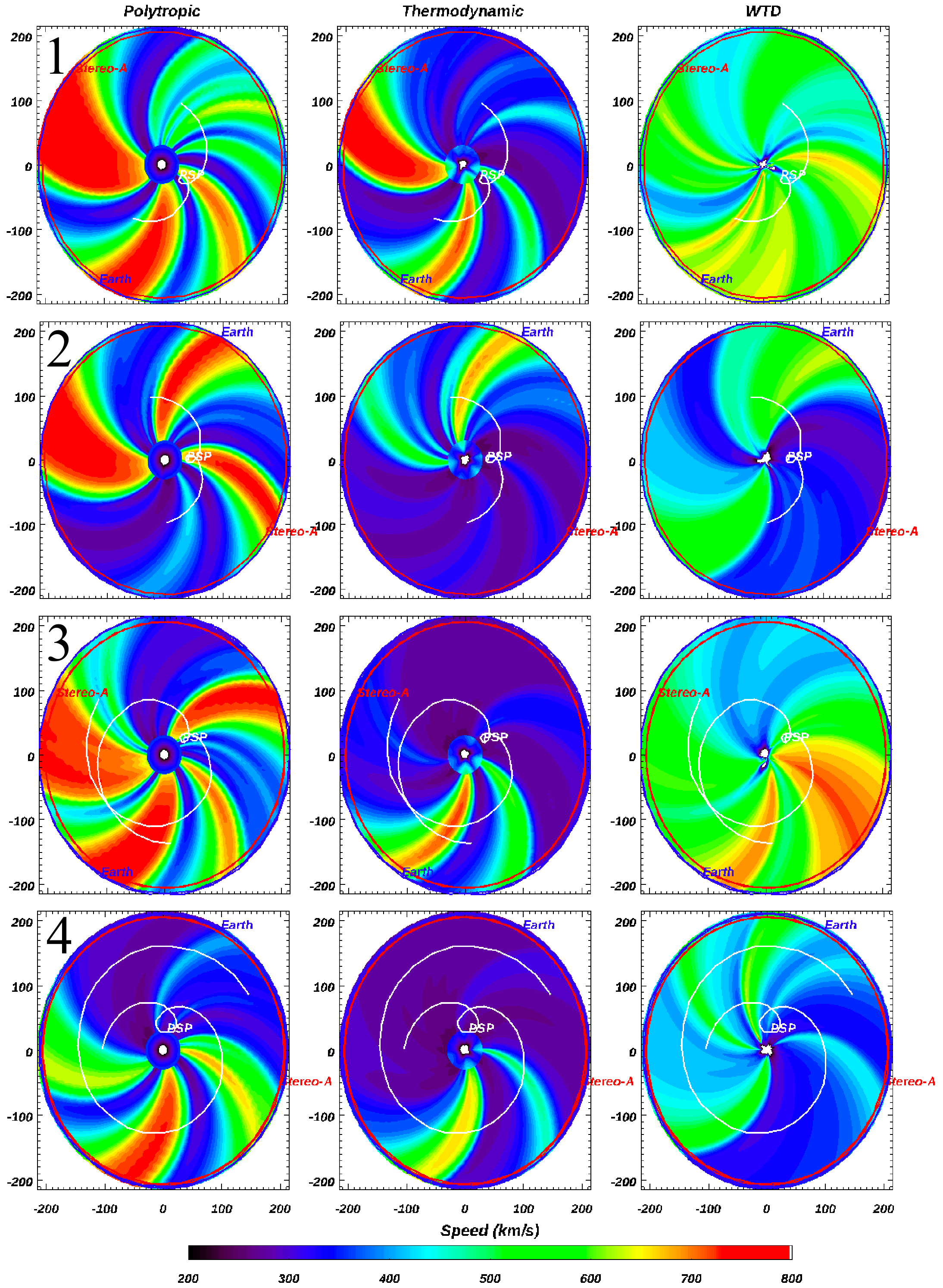}
\caption{
Radial speed profiles in the solar equatorial plane for each model (polytropic, thermodynamic, and WTD) for each perihelion encounter 1-4. Superimposed are the corotating trajectories of the spacecraft during each encounter, expanded by 10 days on either side of the nominal encounter dates. The approximate location of the spacecraft at the time of PSP's point of perihelion pass is indicated by the location of each spacecraft label. 
}
\label{polar-contour-p1234}
\end{figure*}

\subsection{Models}

In this study, we use PSI's MAS (Magnetohydrodynamic Algorithm outside a Sphere) code, which solves the usual set of resistive MHD equations in spherical coordinates on a nonuniform mesh. The details of the model have been described elsewhere (e.g., \citep{mikic94a,riley01a,lionello01a,riley12a,mikic18a,mikic18b,caplan19a}). Here, we restrict our description to several relevant points. 
First, the model is driven by the observed photospheric magnetic field. We use HMI observations from the SDO spacecraft to construct a boundary condition for the radial magnetic field at $1 R_S$ as a function of latitude and longitude. Second, the model is run in two stages: First the region from $1-30 R_S$ is modeled, followed by the region from $30 R_S$ to 1 AU, being driven directly by the results of the coronal calculation. Computationally, this approach is much more efficient. 
Third, for these solutions, we use a single map to cover an entire solar rotation, thus, although the model is time-dependent, it is run forward in time until a dynamic steady-state is achieved. 
This is a reasonable approximation when structure at the Sun is not appreciably changing from one rotation to the next. However, during intervals with significant active region activity, such as E2 and E4 (Figure~\ref{f:magnetograms}), it may lead to inaccuracies. For E2, in particular, the active region was responsible for observed solar impulsive events \citep{pulupa20a}.
Fourth, MAS relies on a variety of approximations to reconstruct (or predict) the large-scale structure and properties of the solar corona and inner heliosphere. In order of increasing complexity (and historical development), we refer to them as the `polytropic', 'thermodynamic' and 'WTD' models. 

The polytropic approximation solves the usual set of MHD equations in spherical coordinates with the energy equation being approximated by a simple adiabatic approximation (that is, we set all energy source terms to zero). This requires us to choose a polytropic index, $\gamma = 1.05$ to reflect the near-isothermal nature of the corona. In the solar wind, it is set to 1.5. These simplifications result in a fast, robust code that reproduces the structure of the magnetic field reasonably well, but fails to generate solutions with sufficient variation in solar wind speeds or densities. To address this, we use an empirically-based approach, DCHB, to specify the solar wind speed at the inner boundary of the heliospheric code \citep{riley01a}. Although semi-empirical, it generally produces results that match 1 AU observations as good as, or better than those computed using the WSA approximation \citep{riley15a}. 

The thermodynamic approximation replaces the polytropic assumption with an empirically-based treatment of energy transport processes (radiation losses, heat flux, and coronal heating) in the corona \citep{lionello01a,lionello09a}. In this case, $\gamma$ now returns to a more defensible value of $\frac{5}{3}$. Development of this model focused on improving the density and temperature structure in the solar corona through comparisons with EUV and X-ray images from a variety of spacecraft. Relatively little direct comparison was performed with in situ measurements, Thus, we also implement the DCHB approximation to derive the heliospheric boundary conditions from the thermodynamic solution. In this sense, the model is not strictly -- or fully -- thermodynamic, and should strictly be labeled the ``semi-empirical thermodynamic'' model. When we have produced complete end-to-end coronal-heliospheric thermodynamic solutions, we have found that they cannot reproduce the observed structure in the solar wind at 1 AU, even though they correctly match the observed amplitude of the quantities. 

Finally, the Wave-Turbulence-Driven (WTD) model is a generalization of the thermodynamic approach by self-consistently heating the corona and using the WKB approximation for wave pressures, providing the necessary acceleration of the solar wind \citep{mikic18b}. The physical motivation for this heating model is that outward and reflecting Alfvén waves interact with one another, resulting in their dissipation, and heating of the corona \citep{zank96a,verdini07a}. And, while the WTD model is significantly more physics-based than, say, the polytropic model, it should be noted that it does require a careful choice of two free parameters. We have found that this approach can account for both the acceleration of solar wind along open field lines, as well as the heating of plasma entrained within closed-field regions \citep{lionello14a,downs16a}; however, again, relatively little exploration of its abilities to reproduce in situ measurements has thus far been performed. 

In summary then, our analysis relies on three models: (1) the polytropic model; (2) the (semi-empirical) thermodynamic model; and (3) the WTD model. Although it is conceptually convenient to think of them as distinct approaches, in reality, they represent snapshots of a continually evolving model; a model that has been developed over $\sim 25$ years and one that has been used to interpret a disparate set of observations at different stages during this evolution. In particular, the polytropic model has been most extensively compared against in situ measurements \citep[e.g.,][]{riley01a,riley12e,riley15a} while the WTD model has been almost exclusively compared against remote solar observations, and, in particular, white-light and EUV images \citep[e.g.,][]{mikic18a,linker19a}. 

\section{Results}

 \begin{figure*}
\centering
\includegraphics[width=17cm]{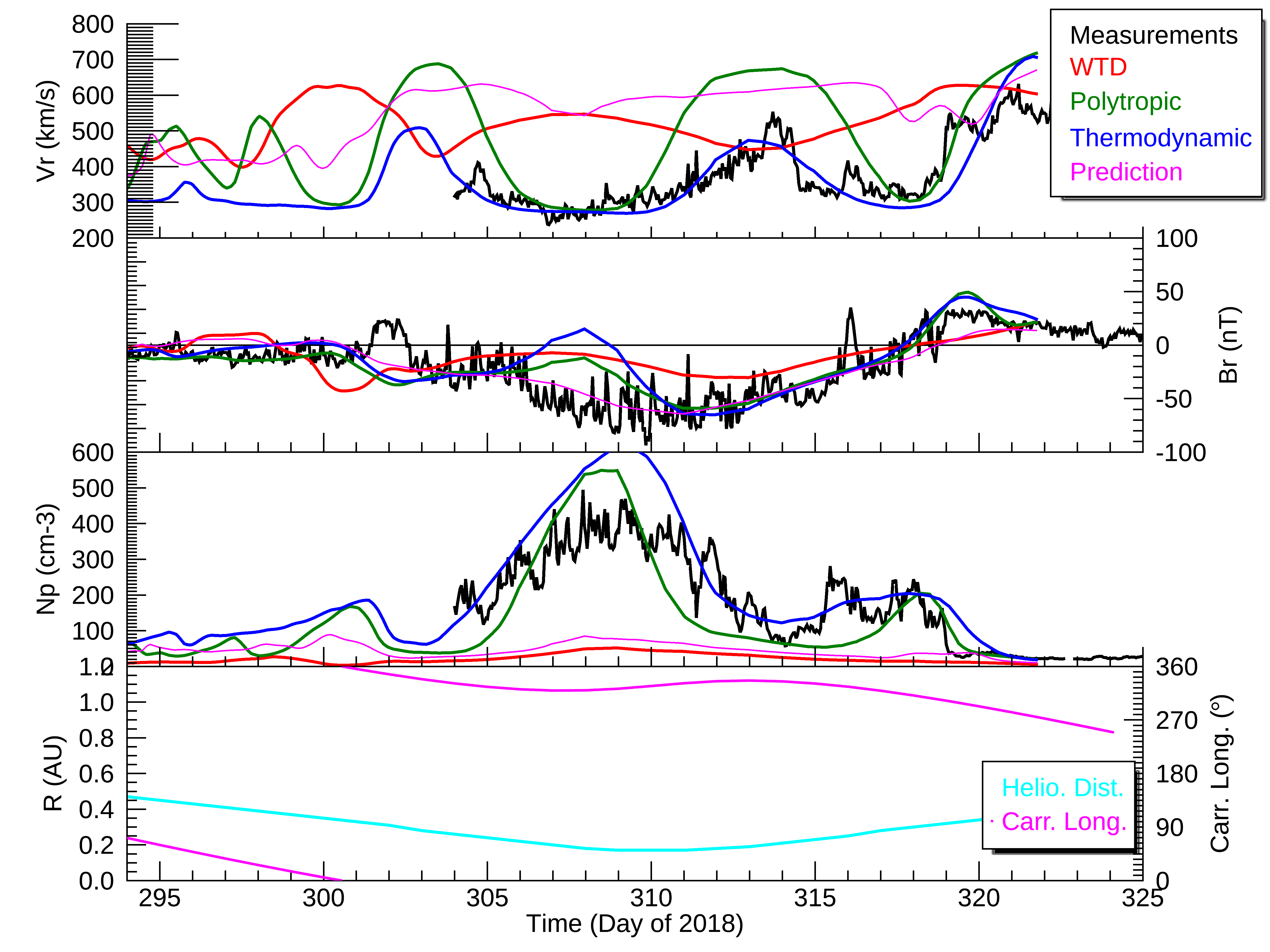}
\caption{
Comparison of model results with PSP in situ measurements during encounter 1.  From top to bottom: Speed, radial magnetic field, and number density ($V_r$, $B_r$, and $N_p$, respectively) are compared with WTD (red), polytropic (green), and thermodynamic (blue) solutions. Additionally, for P1, the original prediction made by \citet{riley19a} is also shown (magenta). Values of $B_r$, with the exception of the prediction, were multiplied by a factor of three. In the bottom panel, heliocentric distance (cyan) and Carrington longitude (magenta) are shown as a function of time. 
}
\label{ts_psp_p1}
\end{figure*}

 \begin{figure*}
\centering
\includegraphics[width=17cm]{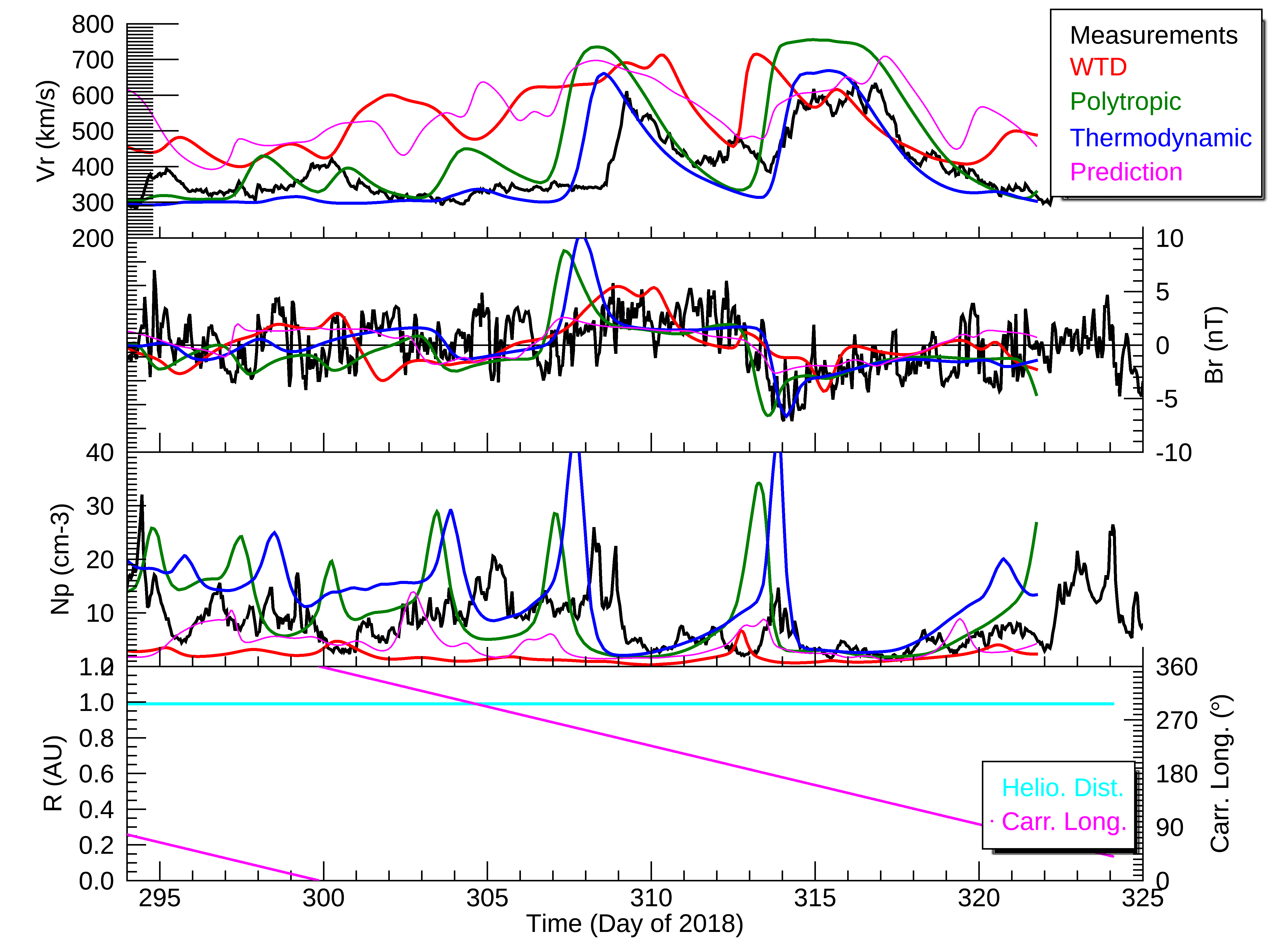}
\caption{
Same as Figure~\ref{ts_psp_p1}, but for model and data P1 comparisons at the location of Earth (with in situ measurements supplied by NASA's SPDF OMNI dataset. 
}
\label{ts_omni_p1}
\end{figure*}

 \begin{figure*}
\centering
\includegraphics[width=17cm]{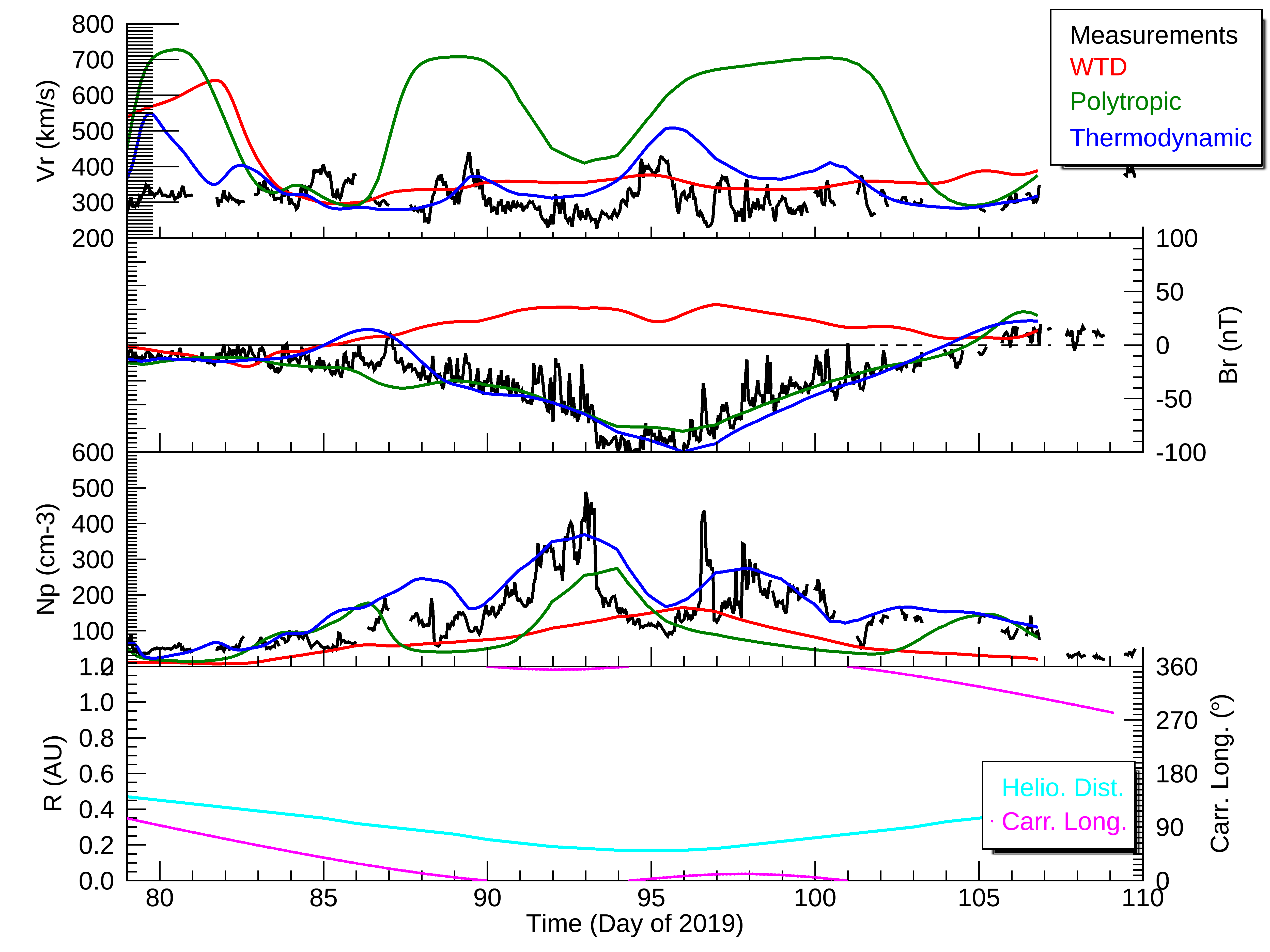}
\caption{
Same as Figure~\ref{ts_psp_p1}, but for PSP data/model comparisons during P2. 
}
\label{ts_psp_p2}
\end{figure*}

\begin{figure*}
\centering
\includegraphics[width=17cm]{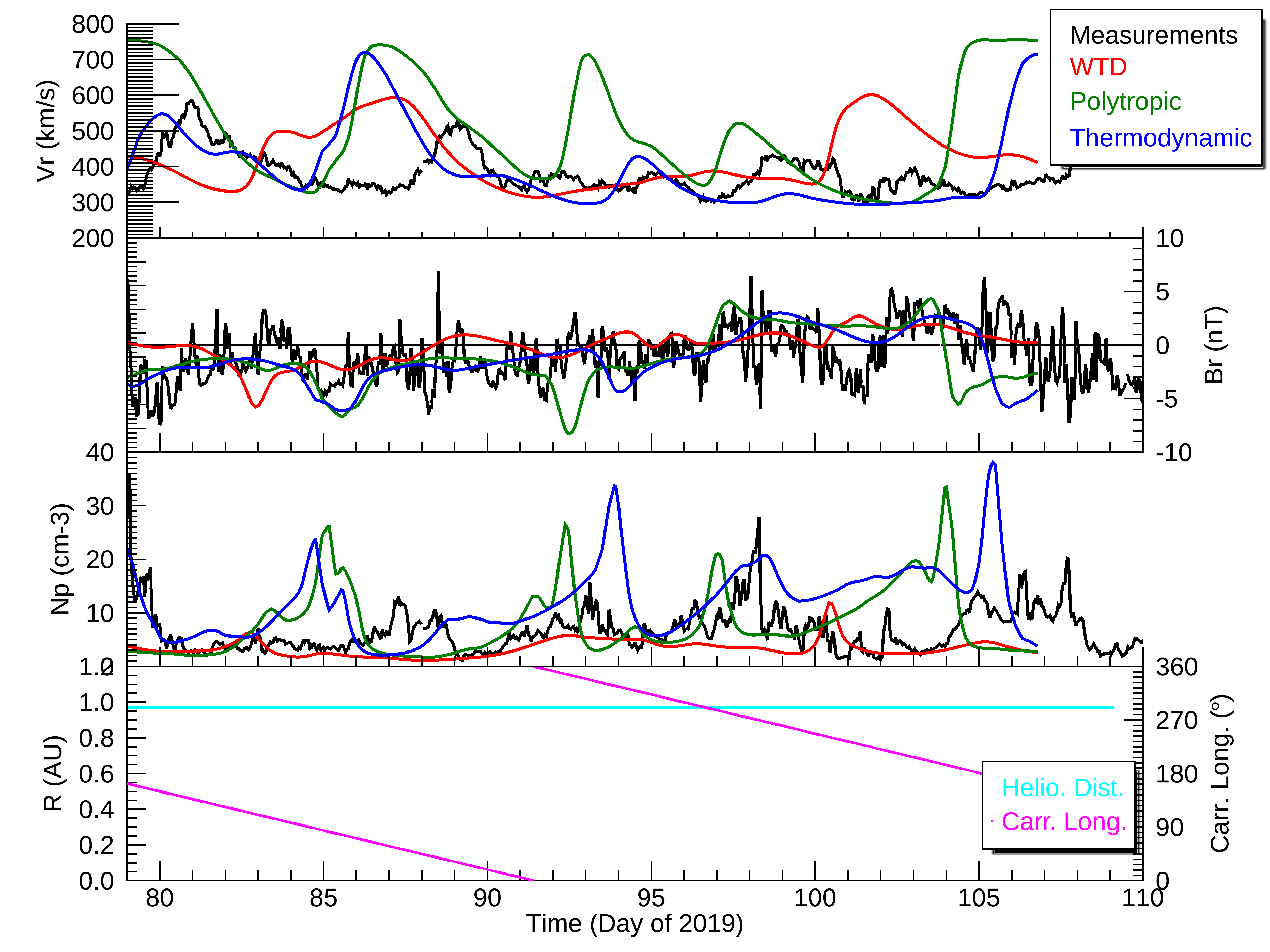}
\caption{
Same as Figure~\ref{ts_psp_p1}, but for model and data comparisons at the location of Stereo-A and for P2. 
}
\label{ts_sta_p2}
\end{figure*}

\begin{figure*}
\centering
\includegraphics[width=17cm]{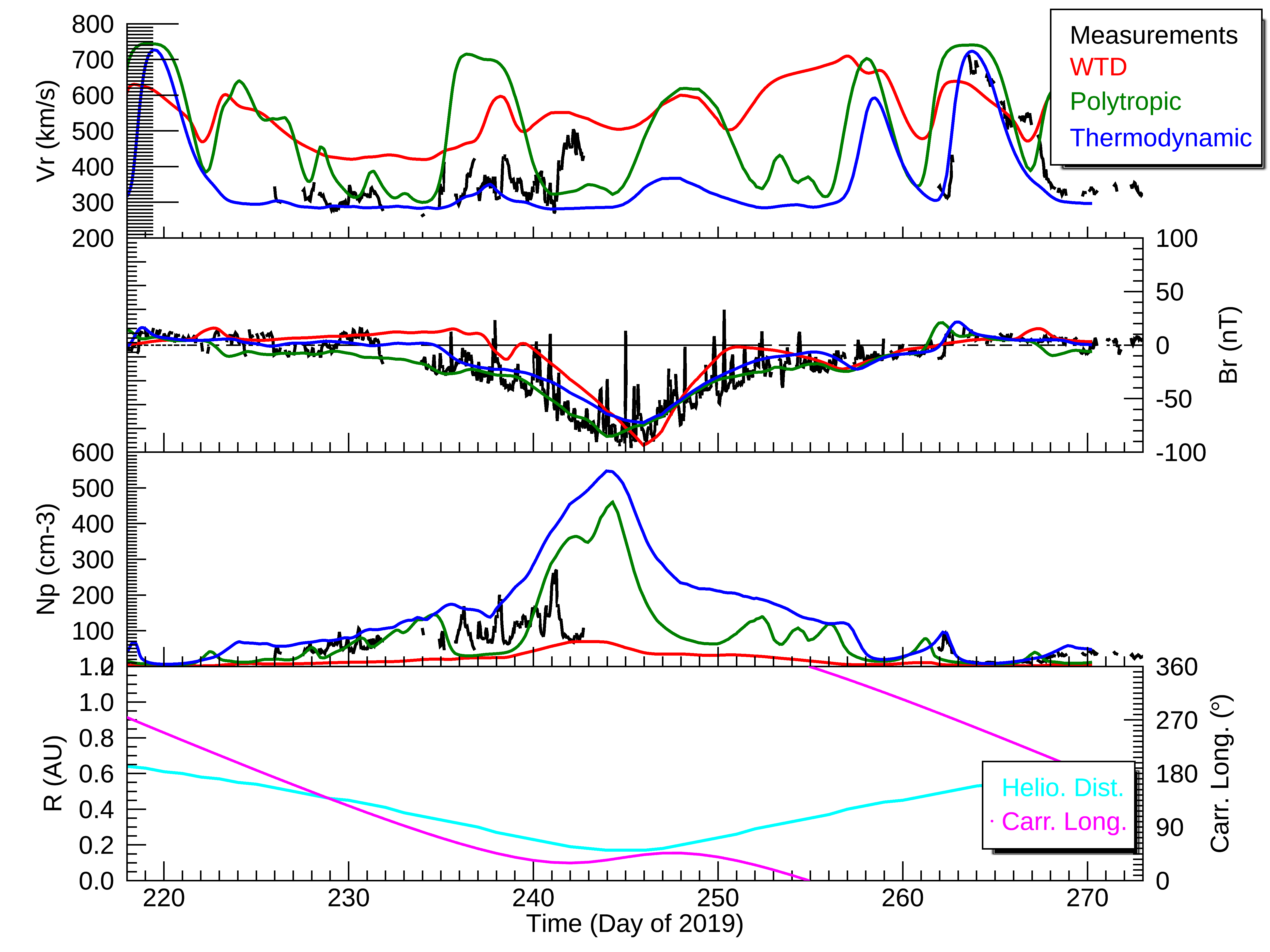}
\caption{
Same as Figure~\ref{ts_psp_p1}, but for model and data comparisons at the location of PSP and for P3. 
}
\label{ts_psp_p3}
\end{figure*}

\begin{figure*}
\centering
\includegraphics[width=17cm]{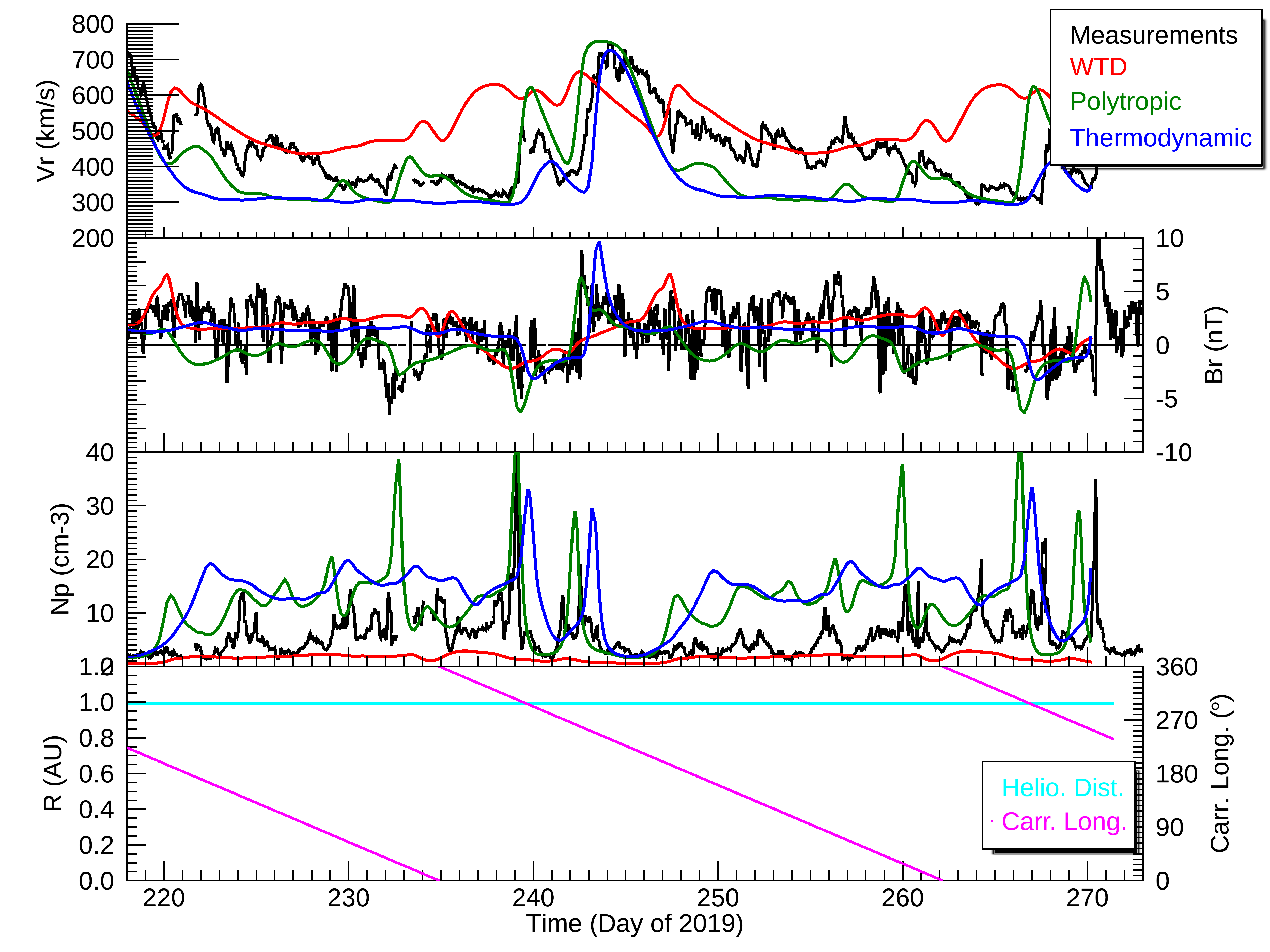}
\caption{
Same as Figure~\ref{ts_psp_p1}, but for model and data comparisons at the location of Earth and for P3. 
}
\label{ts_omni_p3}
\end{figure*}

Before beginning a detailed comparison of the timeseries measurements with the model results, it is instructive to explore the large-scale structural differences predicted by each of the models. Figure~\ref{polar-contour-p1234} summarizes the radial speed of the solar wind from $1 R_S$ to 1 AU for each of the models and for each of the first four PSP encounters (from October 2018 through March 2020). These are displayed in Carrington coordinates with $\phi=0$ along the positive x-axis. Superimposed on each panel is the corotating trajectory of the spacecraft during each encounter, expanded by 10 days on either side of the nominal encounter dates, to encompass the possibility of additional data captured by the spacecraft in some encounters. The approximate location of the spacecraft at the time of PSP's point of perihelion pass is indicated by the location of each spacecraft label. Thus, we can infer that Earth-based observing platforms were able to see the solar origin of the plasma measured at PSP for P4, and, to a lesser extent P2. However, for P1 and P3, the relative position of the spacecraft meant that Earth could not, in particular, provide a direct estimate of the photospheric magnetic field at the time of perihelion. From a modeling perspective then, the boundary conditions used to drive the models for P4 and P2 were intrinsically more accurate than for P1 and P3. We also note how the angular separation of Earth and Stereo-A decreases from P1 to P4. During this interval, Earth was in fact leading Stereo-A; however, this is in fact a consequence of Stereo-A outpacing Earth and lapping it. In progressively later PSP encounters, Stereo-A will again catch (August 2023) and advance past the longitudinal position of Earth. In summary, each encounter presents a modestly different configuration of the three spacecraft. Finally, note that, unlike an inertial projection of spacecraft trajectories, in the corotating frame, the spacecraft travel in a clockwise direction, thus, for example, when interpreting the thermodynamic solution for P1 (top middle panel), PSP measures slow solar wind initially, and, after traversing two modest streams, is immersed in slow solar wind at perihelion (where it hovers at approximately the same Carrington longitude while performing a `loop'), then exits the perihelion as it encounters a significant high-speed stream.  

These maps emphasize the differences in the structure of the solar wind predicted by each model for each encounter. In general, the polytropic model produces stronger high-speed streams in the equator than either of the other two models. Additionally, the WTD model tends to produce broader (in longitude) streams with less contrast between the slower and faster wind streams. Comparing the thermodynamic model to the polytropic model, we note that the structure - as a function of longitude - is quite similar. Based on this, it is likely that the band of solar wind variability for the polytropic model is narrower than the thermodynamic model.

To assess the quality of the model results, we compare data from PSP, Earth-based spacecraft (ACE and Wind, through the OMNI dataset), and Stereo-A for each of the first four perihelia passes (labeled P1 through P4). Beginning with P1, Figure~\ref{ts_psp_p1} compares modeled and observed radial velocity, radial magnetic field, and number density for the interval from day-of-year (DOY) 294 through 325. This includes the nominal 12 days that bracket the date of closest approach (from DOY 304 to 315) but adds an additional 10 days on either side to define an interval that is marginally longer than a solar rotation period. Perihelion occurs near the center of the panel (DOY 310, or 06 November 2020), coinciding with the central portion of the interval, and associated with a positive increase in longitude with respect to time (bottom panel). During the interval from approximately DOY 303 to 316, PSP remained at roughly the same Carrington longitude ($\sim 330^{\circ}$), which, at the time was an unprecedented position for an interplanetary spacecraft to hold. Focusing first on the radial velocity, we note that during this first encounter, PSP was immersed in slow  ($< 500$ km s$^{-1}$ solar wind, only rising substantially beyond this on DOY 319. These variations are well captured by the thermodynamic model results; however, the other models fail to reproduce these basic variations. The WTD model, while reasonably predicting an average and unchanging speed of $\sim 350$ km s$^{-1}$ does not capture any of the structure within the slow stream, nor does it predict the rise in speed later in the interval. The polytropic solution qualitatively matches the variations from slower to faster wind, but the amplitude is far too large, predicting speeds approaching 700 km s$^{-1}$ at the peak of the small stream (DOY 313). Finally, the actual prediction made by \citet{riley19a} is grossly inaccurate (magenta line), with a forecast of constant, 600 km s$^{-1}$ solar wind throughout most of the interval. 

\begin{figure*}
\centering
\includegraphics[width=17cm]{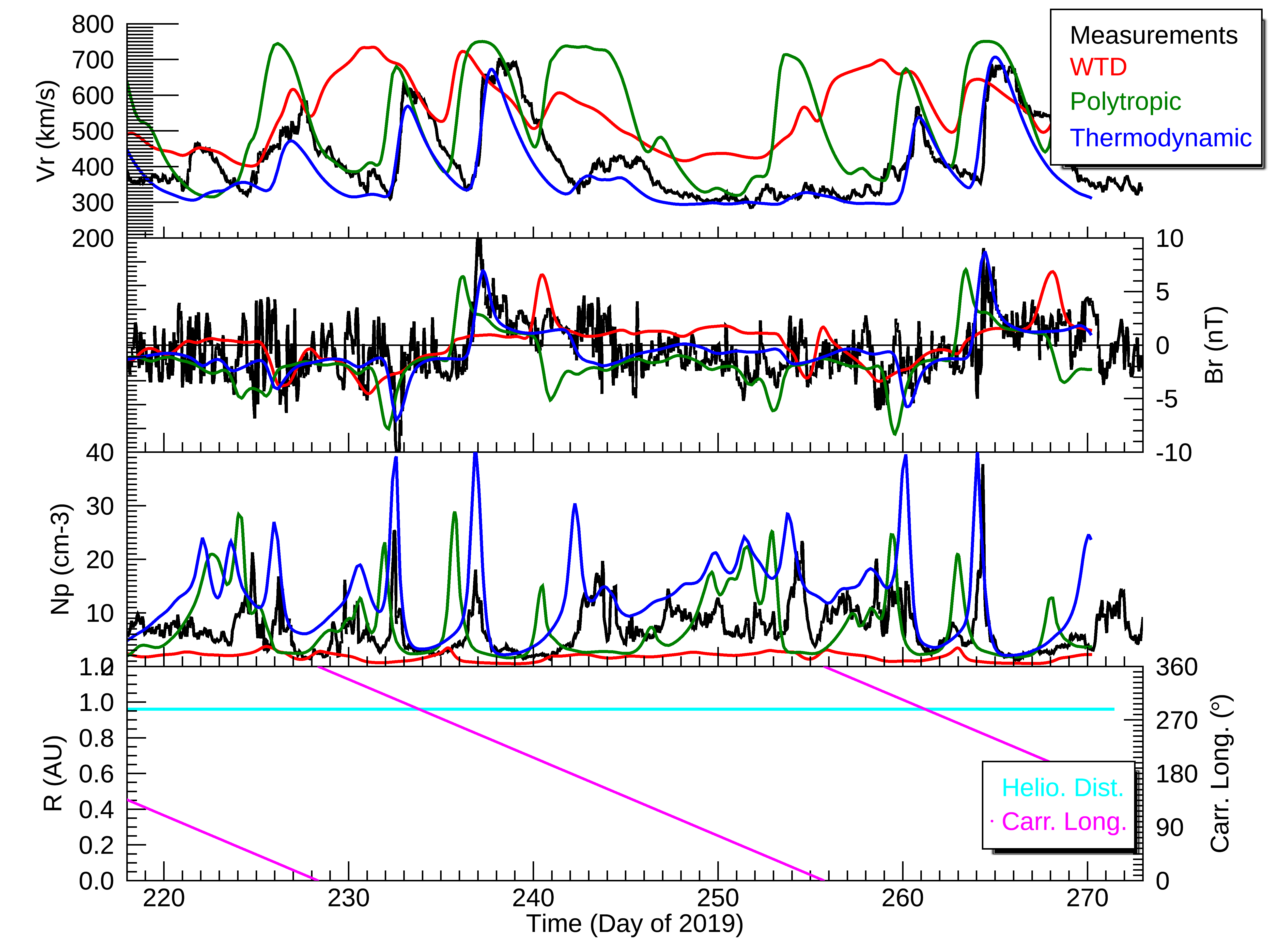}
\caption{
Same as Figure~\ref{ts_psp_p1}, but for model and data comparisons at the location of Stereo-A and for P3. 
}
\label{ts_sta_p3}
\end{figure*}

\begin{figure*}
\centering
\includegraphics[width=17cm]{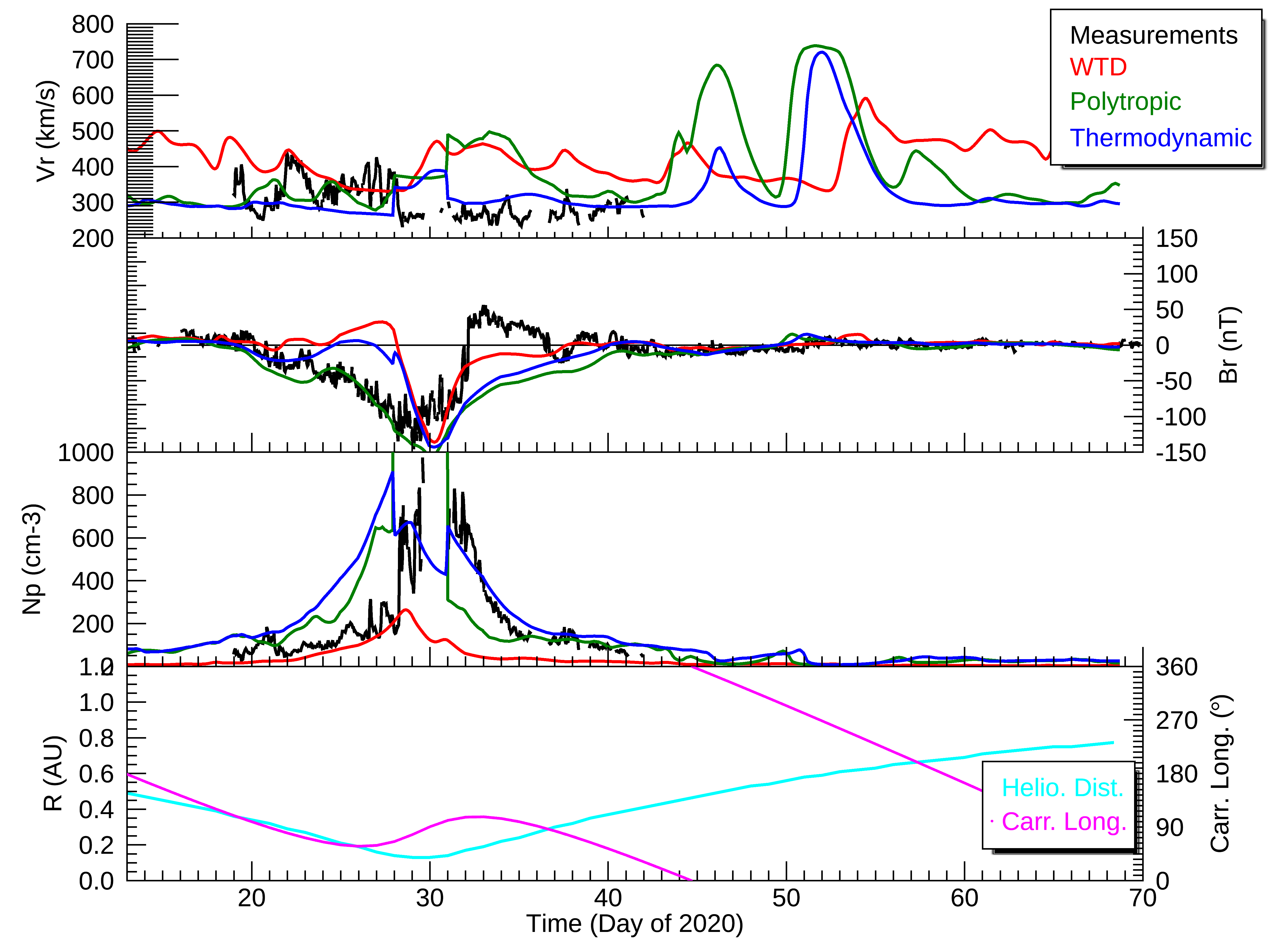}
\caption{
Same as Figure~\ref{ts_psp_p1}, but for model and data comparisons at the location of PSP and for P4. 
}
\label{ts_psp_p4}
\end{figure*}

Similar inferences can be made for the comparisons with the radial magnetic field, $B_r$. It should be emphasized that the values for the polytropic, thermodynamic, and WTD models have been multiplied by a factor of 3, which reflects the well-known problem with global models that, while they are able to capture variations in the field strength, they significantly underestimate their amplitude \citep{riley12e,linker17a,riley19b}. This factor is also the same as used by other global modelers relying on HMI data to drive their simulations \citep[e.g.,][]{holst13a}. Intriguingly, and a point we will return to later, the solution used to make the prediction (magenta) was not multiplied by any corrective factor. 
From these comparisons, we infer that the prediction, polytropic, and thermodynamic solutions appear to have captured the large-scale variations in the magnetic field, and, in particular, the immersion into a negative polarity field for the entire interval, only switching to a positive polarity around DOY 318. Although not shown here, the origin of this wind was an equatorial coronal hole \citep{riley19a,badman20a}. Both the polytropic and thermodynamic solutions suggest some structure in the field around DOY 308 that is not reflected in the observations.  

Finally, we consider density. The third panel of Figure~\ref{ts_psp_p1} compares the model/prediction results with the observed density during the first encounter. Both the prediction and the WTD model fail to capture the substantial increase in density as PSP swept into 35.7 $R_S$. In fact, given that all four model results were approximately the same on DOY 303 and again on DOY 314, this further suggests that the WTD and prediction results were actually regions of lower intrinsic density (a fact confirmed by plots of scaled density, not shown here but presented for the prediction by \citet{riley19a}). Of the two models that capture both the peak at perihelion and the subsequent high-density region following it (DOY 315-319), the thermodynamic results (blue) most closely match the observations, although the maximum value is overestimated. 

The PSP P1 results can be contrasted with Stereo-A and near-Earth spacecraft measurements made at 1 AU. In Figure~\ref{ts_omni_p1} we compare the same three parameters at the location of Earth, using data from the OMNI dataset (composed of measurements from both ACE and Wind). During its passage from the location of PSP to Earth, the streams have evolved, steepening where fast wind is overtaking slower wind ahead, and expanding where slower wind is trailing faster wind. The resulting ``sawtooth'' profile in radial velocity, covering slightly more than a Carrington rotation, is typical of ambient solar wind conditions in the absence of transient phenomena. In this specific case, there are two high-speed streams beginning on DOY 309 and 314. This profile is captured well by both the polytropic (green) and the thermodynamic (blue) models. Of these, the thermodynamic results are again a closer match with observations, mimicking the size and structure of the streams better. Note, in particular, that the first stream has a sharp rise, followed by an immediate decay, while the second one exhibits a flat top before decaying, indicating that the spacecraft was immersed more fully in an equatorial coronal hole (likely the result of an equatorward extension to the polar coronal hole - see later). Again, the WTD results -- both from the prediction and the retrospective run -- do not match the observed structure of the solar wind speed. 

Considering next the radial magnetic field, the observations suggest that the spacecraft skimmed along the heliospheric current sheet (HCS) for most of the rotation, dipping into a positive polarity region around DOY 307 and into a negative polarity region on DOY 314. Only the WTD results miss these sector crossing. We note again, however, that the predicted values have not been corrected by the factor of three that was applied to the WTD, polytropic, and Thermodynamic solutions. 

Comparison of the plasma density with the measurements suggests that the broad variations have been captured by both the polytropic and thermodynamic models. However, both models appear to overestimate the size of the compression regions, driven by the fast streams overtaking slower streams. This is particularly true for the two major streams on DOY 309 and 314. This is likely due to the fact that the speed gradients and overall amplitude predicted by the models were substantially higher than those observed. Additionally, the polytropic solutions systematically predict an earlier arrival of the high-speed streams than was observed. 

Comparison of Figures~\ref{ts_psp_p1} and ~\ref{ts_omni_p1} appears to show that there is more large-scale structure at 1 AU than at $\sim 40 R_S$. However, it is important to note that whereas the same temporal interval at Earth translates into more than 360$^{\circ}$, at PSP, it is approximately half of this (180$^{\circ}$), owing to the spacecraft's acceleration into perihelion and deceleration out of it.

Next we consider P2. Figure~\ref{ts_psp_p2} summarizes the same measurements and model results (with the exception of the prediction results). The solar wind measured by PSP was very quiescent: slow, dense, and somewhat variable plasma emanating from a negative polarity region on the Sun. During the entire interval, and until DOY 105 the field remained negative, thus, it represents a unique opportunity to investigate the properties of solar wind from one specific region, essentially two weeks of measurements came from a Carrington longitude range of less than 20$^{\circ}$ or less, centered near 0$^{\circ}$. In terms of solar wind speed, the thermodynamic and WTD models best match the observed low speed. But, while the actual values are overestimated, only the thermodynamic model appears to have captured the variations, that is, the stream structure. The polytropic model is patently wrong.  From the perspective of the radial magnetic field though, both the thermodynamic and polytropic solutions capture the global features of the field in terms of amplitude and direction. The thermodynamic model erroneously suggests a small polarity reversal starting on DOY 85, but both results are vastly better than the WTD results. Finally, comparing density profiles, once again, the thermodynamic model reproduces the large-scale variations observed in the plasma data, although it somewhat overestimates them. 

Thus far, we have limited ourselves to a qualitative comparison between the model results and the observations. Quantitative estimates, such as mean absolute error (MAE) or correlation coefficient (CC) have their place, particularly for space weather applications, but  can result in inferences that do not match our subjective interpretation \citep[e.g.,][]{owens05a,riley15a,riley13b}, particularly when scientific understanding is the goal. For example, a model that produces a constant velocity at say, 400 km s$^{-1}$, would produce a lower MAE than one that reproduced the structure of the solar wind streams, but not their exact phasing in time.  Clearly the latter model is of more scientific value \citep{riley17b}. To better illustrate this, we computed CCs for each of the model results against their observed values for P2 at PSP (Figure~\ref{ts_psp_p2}). The results were as follows: (1) $CC_{vr}$(WTD,Poly,Thermo): 0.027, -0.197, 0.016; (2) 
$CC_{Br}$(WTD,Poly,Thermo): -0.672, 0.886, 0.8923; and (3) 
$CC_{Np}$(WTD,Poly,Thermo): 0.625, 0.456, 0.804. 
Thus, the quantitative estimates suggest that, in terms of speed, the WTD performs slightly better than the thermodynamic solution. However, practically speaking, based on these results, none of the models shows any significant correlation with the observations. The CCs for the radial field match our qualitative interpretation, with both the polytropic and thermodynamic models outperforming the WTD results. Finally, although our subjective inference that the thermodynamic model substantially outperformed the other two, in terms of density, this is not captured by the CC values, where the CC for the  WTD solution (0.625) does not accurately reflect its lack of agreement in terms of structure, as compared to the thermodynamic solution (CC=0.804). Thus, bearing these points in mind, for the remainder of the study, we continue to provide qualitative descriptions of the model comparisons.

Contrasting the comparison of PSP data/model results with those at Stereo-A (Figure~\ref{ts_sta_p2}) leads to similar statements as for the P1 comparison. Here, it is even more apparent that Stereo-A was skimming along the HCS for the entire interval (as indicated by the radial field values `hugging' a value of zero). When spacecraft are so situated, this makes comparisons with models even more precarious, as small shifts in latitude can lead to substantial changes in the values of the plasma and magnetic field values. The relatively poor comparison across all variables and models reinforces this point. At least qualitatively, one could argue that the thermodynamic solution appears to more closely match the observations; however, this is a weak inference at best. Finally, it is worth noting that the amplitude of the compression regions, seen as the peaks in the number density profiles are largest for the thermodynamic model. This may appear paradoxical because the differences in speed between the slow and fast streams are larger for the polytropic results than the thermodynamic results. However, the reason is that the base density in the slow solar wind of the thermodynamic model is substantially higher. Thus, the compression of an already denser medium by a relatively smaller high-speed stream leads to a larger compression region (as measured by peak density) than the compression of a more tenuous region, even if the fast wind compressing it is substantially faster. This is reinforced by comparing with Figure~\ref{ts_omni_p1}, which shows similar effects. In general, the thermodynamic compression regions are significantly larger than are observed, suggesting that the model could be improved by setting the base number density for the slow wind, which is a model parameter, to a lower value. 

Turning next to P3, in  Figure~\ref{ts_psp_p3} we compare radial speed, radial magnetic field, and number density once again with the three model results. Unfortunately, during this perihelion some of the plasma measurements were unavailable. Nevertheless, what was recovered, again, demonstrates that the thermodynamic solutions most closely match the data, although again, the base number density is too high. Note, in particular, the generally low speed prior to the perihelion, and the large high-speed stream on DOY 63, which the thermodynamic model captures well both in terms of phase and amplitude. This, comparison, in turn, suggests a potentially interesting application for the models: The thermodynamic solution may be a reasonable proxy for the unobserved data during the actual perihelion portion of the mission. This is also supported by the reasonable match between the model results and the observations of the radial component of the magnetic field shown in the second panel, as well as comparisons with data at 1 AU, which we turn to next.

\begin{figure*}
\centering
\includegraphics[width=17cm]{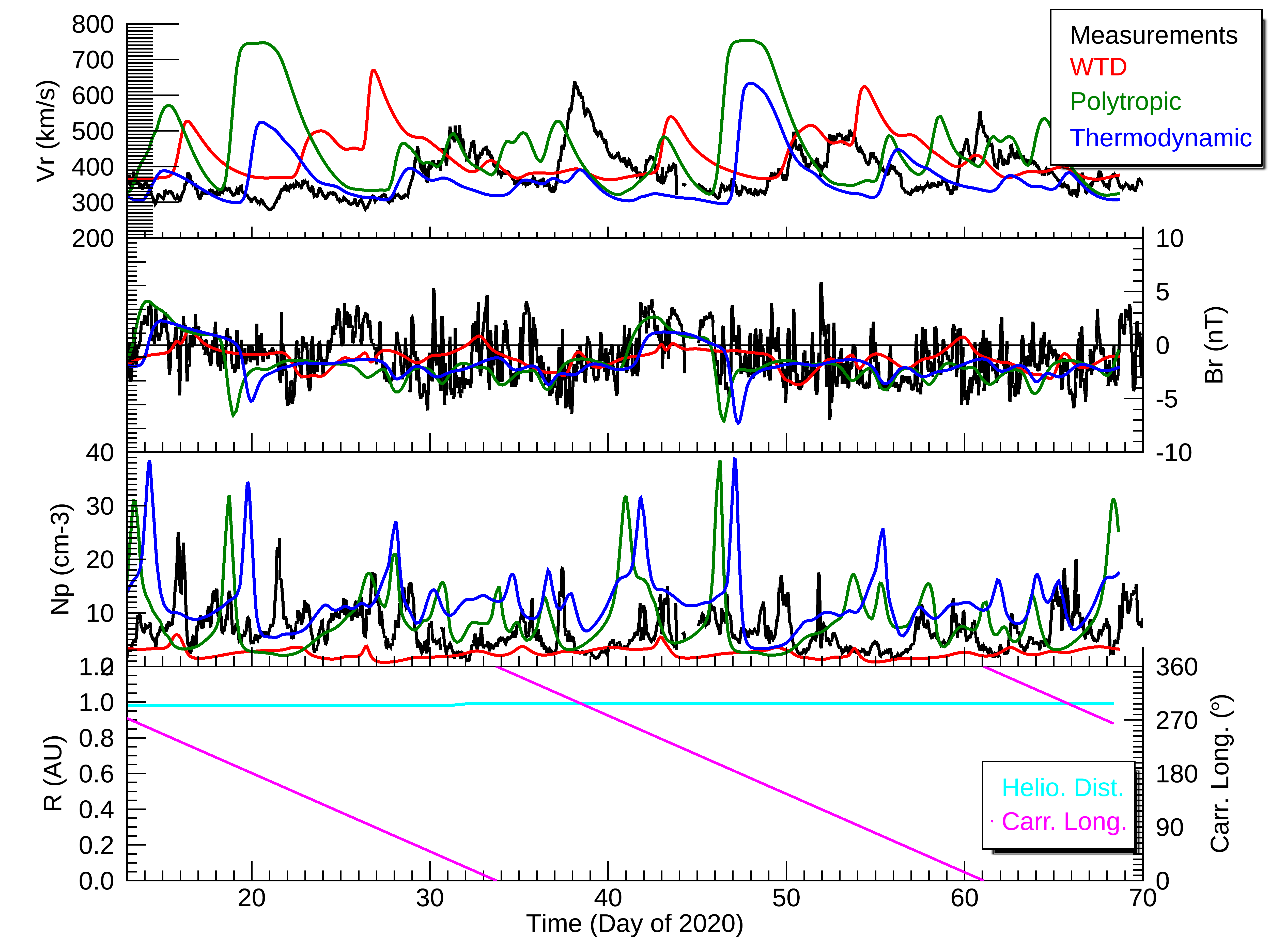}
\caption{
Same as Figure~\ref{ts_psp_p1}, but for model and data comparisons at the location of Earth and for P4. 
}
\label{ts_omni_p4}
\end{figure*}

Perihelion 3 at Earth looked quite different (Figure~\ref{ts_omni_p3}). Although there was a single large stream of $\sim 700$ km s$^{-1}$ this occurred some 20 days prior to the arrival of the stream at PSP. Although it is tempting to associate this stream with the single high-speed stream observed at PSP, inspection of Figure~\ref{polar-contour-p1234} shows that, at least in the equatorial plane, there were two possible candidates for the stream. To assess this more carefully, we note that at the time of PSP's perihelion, Earth was located at approximately $232^{\circ}$ Carrington longitude, which coincided with the arrival of the second and more significant high-speed stream. Additionally, referring back to Figure~\ref{ts_psp_p3}, we can infer that the observed high-speed stream at PSP was probably the second of the two, based on the thermodynamic profile showing two peaks, with the second one being more pronounced and aligning well with the measurements. Unfortunately, we cannot be certain because of the data gap. Nevertheless, based on the global perspective provided by the model, it is reasonable to infer that the high-speed stream observed at PSP on DOY 263 was the same one observed at Earth on DOY 243, one solar rotation earlier. 

More generally, we can make several comments about the comparisons between models and observations. First, although the single high-speed stream is produced by both the polytropic and thermodynamic models, little of the structure between this and the next stream on DOY 268 is captured. Similarly, only some of the structure during the first half of the interval matches. On the other hand, the two-sector pattern shown in the radial magnetic field is reproduced, with the spacecraft remaining predominantly on the positive side of the HCS. All models, to varying degrees, capture this sector structure, with the thermodynamic model performing best. Finally, the modeled densities during this interval are systematically either too high (polytropic and thermodynamic) or too low (WTD), which is related to the fact that the modeled speeds are systematically too low or high, respectively. 

The comparison between model results for P3 at Stereo-A is considerably better (Figure~\ref{ts_sta_p3}). There were five/six high-speed streams of various strengths during this two-rotation interval, almost all of which were captured by the thermodynamic model. This is reinforced by the density measurements, for which the compression regions (peaks in density) match with the observations, albeit in most cases being too strong. The polytropic and WTD solutions did not reproduce this structure with any degree of fidelity, and, as with the Earth comparison, tended to over-estimate or under-estimate the densities. Unfortunately, magnetic field data was only available for a portion of this interval. Nevertheless, the thermodynamic solution, in particular, matches most of the large-scale variations, and, in particular, the transition from a toward to an away sector on DOY 237.  Based on the close match between the thermodynamic solution and the plasma measurements at Stereo-A, it would not be unreasonable to infer the unobserved large-scale magnetic structure from the thermodynamic model results, and, in particular, the transition from negative to positive polarity occurring rapidly and strongly on DOY 264. A final point that is worth reinforcing is that, again, while the properties of the plasma speeds are remarkably well captured by the thermodynamic model, the modeled density peaks in the compressions are significantly larger than were observed. This, again, we believe is a result of setting the boundary value for the slow solar wind to be larger than was observed, providing more material for the high-speed streams to compress.  

The final PSP encounter analyzed here, P4, occurred during the first two months of 2020. Figure~\ref{ts_psp_p4} contrasts the model results with the observations. In general, we remark that the overall features observed in the observations are matched by the thermodynamic solutions. These include: (1) generally slow (300-400 km s$^{-1}$) solar wind; (2) peak densities, likely exceeding 800 cm$^{-3}$; and (3) peak radial fields reaching almost -150 nT. If the model results are an accurate predictor of what PSP would have observed later, we might predict the appearance of a high-speed stream (700 km s$^{-1}$) on DOY 51. However, given the limited data and discrepancies, this would be a tentative conclusion at best. These discrepancies include: (1) the earlier rise in number density in both the thermodynamic and polytropic solutions than was observed; (2) the failure to capture the switch from negative to positive magnetic polarity on DOY 32; and (3) an apparent discontinuous change in the speed and density of the polytropic and thermodynamic solutions close to perihelion. This last disagreement reveals an interesting limitation of the models, only appearing because of the distance of closest approach during P4. For our standard web-based model solutions, the boundary separating the coronal and heliospheric models was set to $30 R_S$. For the previous encounters, this did not impact the simulated spacecraft fly-throughs since they always occurred within the domain of the heliospheric model. For P4, however, PSP's closest approach was 26.9 $R_S$, which is inside this boundary. Since both the polytropic and thermodynamic models are not seamless at this boundary, artifacts are introduced when the simulated spacecraft is flown through the merged solution. This is particularly noticeable for the polytropic solution, for which the coronal model cannot adequately calculate the speeds and densities of the plasma. And, while the thermodynamic solution is better, because we did not drive the heliospheric model directly with the coronal solutions, discontinuities are inevitably introduced. Accepting these limitations, however, we remark that, again, the thermodynamic model appears to estimate the densities and speeds best. For the magnetic field, such discontinuities are not as significant, since the field at the outer boundary of the coronal solution is used directly to drive the inner boundary of the heliospheric model. 

 \begin{figure*}
\centering
\includegraphics[width=18cm]{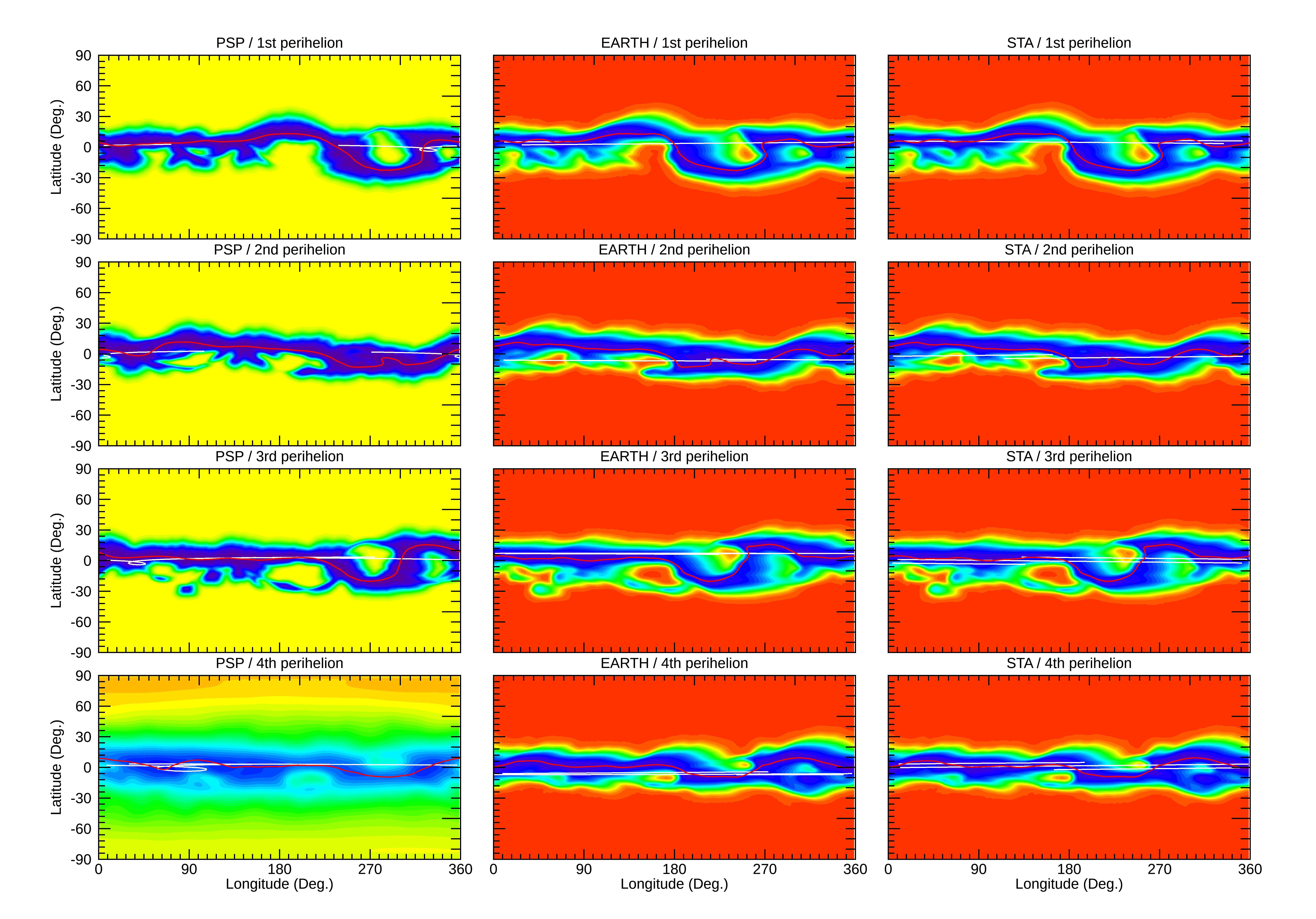}
\caption{
Comparison of MHD modeled solar wind speeds at the location of the three spacecraft (PSP, Earth, and Stereo-A (STA)) with the interval defined by each PSP perihelia pass (1-4).  In each panel, speeds from 200 to 800 km s$^{-1}$ are shown in the rainbow spectrum from blue to red, using the same transitions as in Figure~\ref{polar-contour-p1234}'s colour bar. The orbit of each spacecraft is shown by the white curve and the location of the HCS is marked by the red contour ($B_r = 0$). 
}
\label{vr-lonlat-matrix}
\end{figure*}

As a final comparison, in Figure~\ref{ts_omni_p4} we compare in situ measurements with model results at the location of Earth. This is more representative of model comparisons in the ecliptic plane, particularly when there are not any strong equatorial sources of high-speed wind and/or the spacecraft is skirting close to the HCS. In such cases, an argument can be made that some of the stream structure is captured, but that the phasing of the streams in time is not precise. This, however, is a subjective interpretation, requiring support from other data sources. One such dataset is the magnetic field, which, in this case, matches reasonably well: most of the interval is associated with negative polarity field. An excursion into positive polarity field on DOY 42 is well matched by the thermodynamic and polytropic solutions, as well as the initial descent from positive to negative polarity at the start of the interval; However, another potentially significant excursion into positive polarity on DOY 25 is completely missed by all models. Comparisons with the number density highlight the mismatch in phasing of the modest compression regions and the overestimates of the base densities for both the thermodynamic and (to a lesser extent) polytropic solutions and the underestimate by the WTD model. Overall, this comparison does not provide compelling support that any of the models have reproduced the large scale structure of the heliosphere for this interval. On the other hand, we recognize that the spacecraft's proximity to the HCS suggests that the model results would necessarily introduce large uncertainties.

In summary, this detailed comparison between several models and observations at PSP, Stereo-A, and Earth for each of the first four perihelia passes, provides a means to evaluate the model results in terms of our confidence in their ability to reproduce the global structure of the heliosphere. The results suggest that, in general, the 'thermodynamic' approach reproduces the observations most closely, and thus, we advocate that people use this model from PSI's website. Additionally, it supports the view that the underestimate of the magnetic field by the models is a feature that persists at least to within 26.9 $R_S$, and, thus, cannot be resolved by interplanetary processes. Finally, these comparisons suggest that the model results are most accurate for P1 and P3.

Based on these comparisons, we can now investigate the global structure of the inner heliosphere predicted by these models. Given the particularly good comparisons for P1 and P3, we infer the accuracy of these results is highest, but that, likely, the large-scale picture drawn from the models for P2 and P4 is also reasonable, although, subject to the caveats noted earlier about the presence and likely short-time-scale evolution of the active regions during these intervals. Also, since the thermodynamic solutions consistently performed better than the other two model approaches we limit our analyses to these results.  

Figure~\ref{vr-lonlat-matrix} summarizes the radial velocity profiles from the thermodynamic model for each of the four perihelia at each of the spacecraft. Note that the maps for Earth and STA are virtually the same, since they are both close to 1 AU. The latitudinal position of the spacecraft, however, can be quite different, leading to the striking differences in the in situ comparisons. Since PSP was also moving in heliocentric distance as well as longitude and latitude, this slice is taken from the distance of closest approach; thus, it is most representative of the point of perihelion, which is indicated by the loop in the white trajectory curve. Perhaps the most visually significant difference between the maps is the slower speeds seen at PSP, reflecting the fact that the solar wind is continuing to expand between PSP and 1 AU. Several other features are worth noting. First, between PSP and Earth (or STA) there is a general increase in the complexity of the structure. Whereas at PSP, there is a simpler two-state picture of slow and fast wind, at Earth (or STA), the fast wind within the latitudinal bands of the HCS has become more isolated, almost forming islands (or beams in the radial sense) of fast wind. Second, the structure of the overall band of solar wind variability \citep{gosling95d}, that is to say, the region within $\pm 30^{\circ}$ latitude has increased: Whereas at PSP it was relatively flat, by 1 AU it arcs substantially more in latitude. Third, the latitudinal gradient in solar wind speed at PSP has disappeared and wind beyond this band of solar wind variability is approximately 750 km s$^{-1}$. Fourth, the underlying reason for the differences in the timeseries profiles seen in Figures~\ref{ts_psp_p1} - \ref{ts_omni_p4} between the three spacecraft can be, at least in part, interpreted as differences in the latitudinal position of the spacecraft at each of these times. In fact, generally speaking, Earth and PSP/STA were always separated the most in latitude, with the former generally in the southern hemisphere and the latter in the northern hemisphere. Moreover, this suggests that PSP and STA should have much more similar profiles, at least based on the relatively similar latitudinal position. As an example, comparison of PSP and Earth for  P3 reveals that PSP was immersed in slow solar wind for most of the time, and particularly at perihelion, whereas Earth partially intercepted a high speed stream, but was generally embedded within solar wind of more variable speed. 

Finally, comparison between PSP and either Earth or STA maps, particularly for the first and third perihelia, highlights the evolution of stream structure in moving outward from the location of PSP (a few tens of solar radii) and Earth (215 $R_S$). Note in particular, how the HCS (red trace) becomes distorted, with regions at highest latitude getting pulled to earlier longitudes in some locations and stretched to later longitudes at other locations. This is due to the dynamical effects of the surrounding stream structure. Fast solar wind at earlier longitudes (smaller heliocentric distances) attempts to overtake slower wind at later longitudes (farther heliocentric distances), decelerating the slower wind within which the HCS is embedded. Similarly, when fast solar wind outruns slower solar wind behind, the current sheet is stretched out in longitude (as it is embedded within the expansion wave (rarefaction region) that is created through this process). 

\begin{figure*}
\centering
\includegraphics[width=18cm]{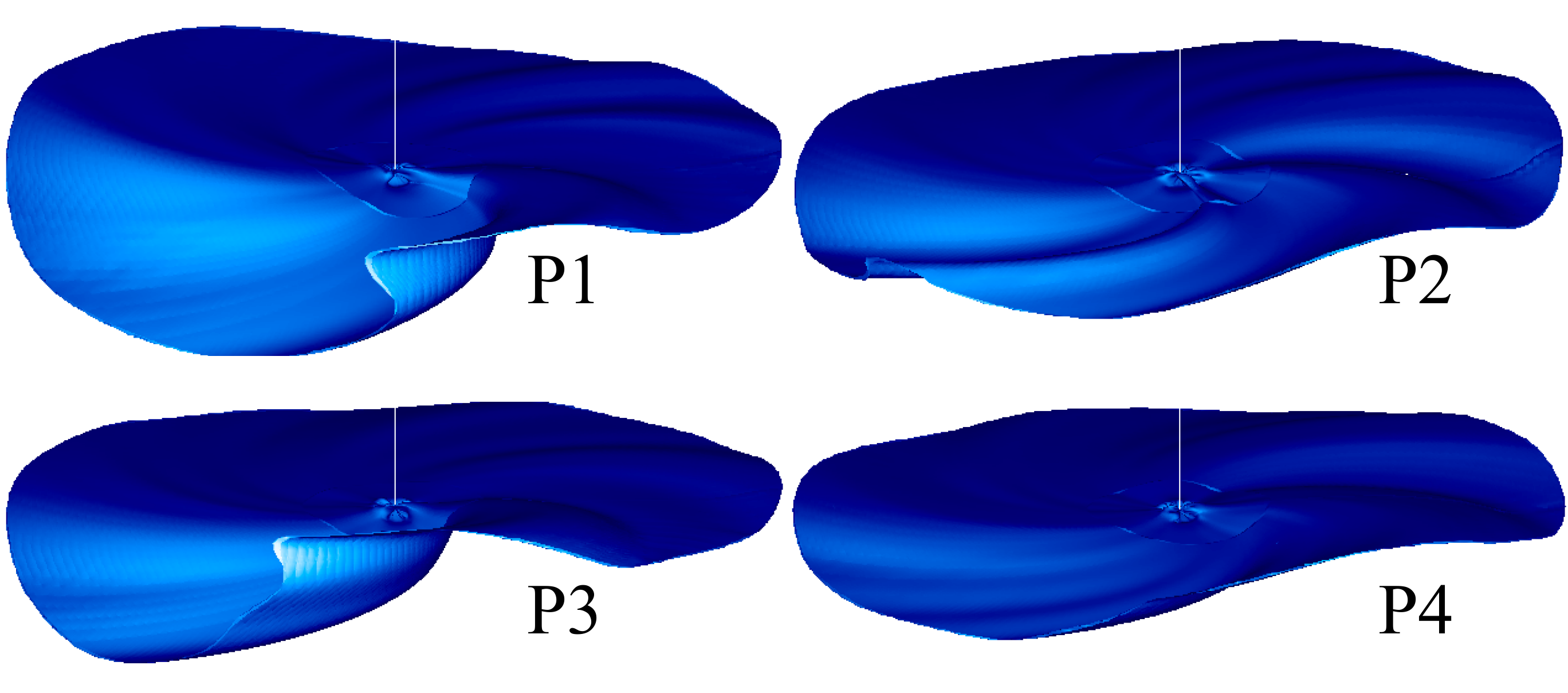}
\caption{Comparison of the shape of the HCS for each of the first four perihelia passes (P1, P2, P3, and P4). The HCS is approximated by the isosurface of $B_r = 0$ from the solar surface to 1 AU.  The thermodynamic solutions were used to estimate the location of the HCS for each perihelia, and the slight discontinuity visible at $30 R_S$ is due to the merging of the coronal and heliospheric solutions. 
}
\label{hcs}
\end{figure*}

The HCS is effectively a surface separating regions of opposite magnetic polarity, and, although not directly observable, it is the largest coherent structure within the heliosphere. It is intimately related to the large-scale dynamical flow of the solar wind, and although passive, it responds to the dynamics of interplanetary streams and thus provides an indirect measure of stream evolution. At least during relatively quiescent times (and within, say, 20 AU), corotating interaction regions (CIRs) are organized about the HCS \citep{pizzo94a}. The large-scale structure of the HCS is summarized in Figure~\ref{hcs} for each of the four perihelia passes. In general, we note the relatively flat orientation of the HCS for all four perihelia passes. These pictures can be contrasted with the shape of the HCS during more active conditions, which can span effectively the entire spherical domain \citep{riley02a}.  More specifically, however, we note that for P1 and P3, there was a vertical, and hence sharp crossing of the HCS from negative polarity (southern hemisphere) to positive polarity (northern hemisphere), whereas for perihelia P2 and P4, any crossings of the HCS were at more inclined angles. Additionally, the HCS is notably flatter during the interval surrounding P4. These are consistent with the inferences drawn from Figure~\ref{vr-lonlat-matrix}, and explain why the model comparisons with observations were generally better for P1 and P3, for which there were well-defined crossing of the HCS and immersion into distinct sources of solar wind on either side.

\section{Conclusions and discussion}

In this study we have modeled the global structure of the inner heliosphere for each of the first four PSP perihelia. We found that our semi-empirical thermodynamic model consistently produced better matches with observations than either a less sophisticated polytropic approach, or a more advanced wave-turbulence-driven model. Our results provided a global perspective from which to interpret the localized in situ measurements made during P1 through P4 and connect observations made at PSP with those at 1 AU from Earth-based spacecraft as well as Stereo-A. We did not find any evidence for the resolution of the open flux problem at least within the heliocentric distance reached by PSP during P4 ($\sim 26.9 R_S$). 

While the thermodynamic model results were, in some cases, quite remarkable, the agreement between model output and observations may be improved upon in a number of ways. First, and foremost, and as described previously \citep[e.g.,][]{riley12e,linker17a,riley19b,riley19a}, the model results are extremely sensitive to the boundary conditions, and, in particular, the observed photospheric magnetic field. The limitations of these measurements include: (1) No observations beyond the view afforded by Earth-based spacecraft, which includes limited inferences on the evolving field on the backside of the Sun as well as little-to-no reliable information from polar latitudes; (2) No consensus on the actual ``ground truth'' values of the photospheric field \citep{riley14c}; and (3) Limited availability or accuracy of nonradial magnetic fields at the base of the model. 

Second, and related to this first point, is the lack of availability of accurate time-dependent boundary conditions.Currently, flux-transport models, such as ADAPT \citep{arge10a}, are the likely the best quasi-time-dependent solution to this problem; however, they are driven by Earth-based observations of the photosphere. Only with the availability of non-Earth views of the Sun, or the accurate assimilation far-side data from helioseismic techniques \citep[e.g.,][]{liewer17a}, can we begin to build truly time-dependent, synchronic maps of the Sun. With the successful launch of Solar Orbiter and the availability of observations away from the Sun-Earth line from the polarimetric and helioseismic imager (PHI) \citep{solanki14a}, we can begin to assemble boundary conditions that mitigate and quantitatively assess the impact of this limitation.
Additionally, PFSS solutions, particularly for PSP encounters is reinforcing the idea that even daily updated quasi-synchronic maps, such as produced by ADAPT (using either GONG or HMI observations) can produce significantly better matches in terms of the observed polarity of the magnetic field measured by FIELDS \citep{badman20a}. Thus, a natural but challenging next step would be to assess how driving the MHD solutions with a sequence of ADAPT maps affects the quality and accuracy of the solutions.  

Third, the free parameters set in each of the three models have not been rigorously tested in sensitivity studies. Over the years, we have explored heuristically how different values might impact specific comparisons with data (e.g., white-light, EUV/x-ray, in situ); however, no systematic study has been performed. Moreover, these values were, in some cases, set during periods of solar activity that was substantially different to the current state of the corona. For example, the free parameters used to specify the mapping of solar wind speed along field lines in the corona, to drive the heliospheric boundary conditions of the polytropic and thermodynamic models were essentially fixed based on specific (but detailed) analysis of the time period surrounding the solar minimum of 1996 \citep{riley01a}. Although they remain reasonable, based on the comparisons presented here, it is likely that the change in the overall state of the corona during that past quarter of a century might require a re-examination of these parameters. 

Fourth, and finally, model comparisons with observations may be improved by incorporating more datasets into the assessment of the model results, as well as defining and using more quantitative metrics for accessing accuracy. Currently, we use an ad hoc approach of comparing subsets of the available data, depending on the specific datasets we are hoping to interpret with the model results. For example, in eclipse predictions \citep{mikic18a}, the focus is to produce the best match with white-light observations of the actual eclipse. This is, however, at the expense of matching in situ measurements. We have suggested that an approach that incorporates metrics for all available metrics, within the framework of a Pareto frontier \citep{camporeale20a}, may optimize model solutions. Of course, it could be that the resulting solutions are neither the best ones at reproducing coronal observations nor in situ measurements. 

It is noteworthy that, overall, the thermodynamic model appears to be outperforming both the polytropic and WTD approaches. The polytropic model represents the most empirically-based technique, while the WTD model is the most physics-based. For many years, the polytropic model performed best in comparisons with in situ measurements over a wide range of intervals covering the space era (e.g., \citet{riley01a,riley12b,riley12e}. The motivation for the thermodynamic model was, at least in part, to address the limitation that the polytropic approximation resulted in poor comparisons with white-light observations. It is not yet clear whether the improved in situ results of the thermodynamic solutions are due to the overall improvement of parameters in the model controlling the heating of the corona, the new lower-level of solar activity that the Sun has entered, or some other phenomena. Either way, it is encouraging that a more sophisticated model is now capable of providing more accurate solutions. Of course this requires further, more systematic comparisons with observations over a broader range of the solar cycle. 

The primary distinction between the polytropic and thermodynamic solutions originates in the structure of the magnetic field in the corona, since both heliospheric solutions are driven by this. Thus, the fact that the thermodynamic solutions are providing closer matches with in situ measurements reinforces previous coronal comparisons with white light observations, particularly during eclipses, which also demonstrated the superiority of the thermodynamic results. A further refinement to the thermodynamic approach could be made, bridging the gap with the WTD model, by directly driving the heliospheric model with the output from the thermodynamic solution. However, previous explorations of this have resulted in worse comparisons with 1 AU measurements \citep{riley12e}. 

The investigation undertaken here was an exploration of several different modes of operation of several models, to understand, at least qualitatively, the capabilities of the different approaches. It should not be interpreted as a systematic parameter phase-space of sensitivity study, which would require a more careful analysis, changing only one variable at a time and assessing the impact of that change on the model results and their similarities and differences with observations. For example, in comparing the polytropic/thermodynamic solutions with the WTD approach, the biggest distinguishing factor might be that the maps were processed using different approximations and had little to do with the underlying physics in the model. We found that the updated pipeline for the WTD solutions resulted in better comparisons with white-light observations during the last two total eclipses in 2017 \citep{mikic18a} and 2019 \citep{linker19a}, and we had anticipated that this would translate into better comparisons with in situ measurements. That it didn't, now requires another suite of runs where we use the same input maps for all modeling approaches. 

Intriguingly, our study showed that, in spite of our initial prediction for PSP's first perihelion, there remains a significant deficit in the value of the radial (and hence total) magnetic field predicted by the models. Our prediction, which relied on older photospheric magnetic field data, likely estimated the value of $B_r$ poorly, at least in part due to the significant evolution of the active region located at PSP's subsolar point as it reached closest approach. Unfortunately, this conspired in just the right way so that when the measurements were returned from the spacecraft, the agreement appeared to be remarkable. Although this hinted at the possibility that the so-called open flux problem might arise in processes in the solar wind (beyond several tens of solar radii), it is clear from the results presented here (requiring a factor of 3 correction to bring the model results into agreement with the observations) that the problem exists throughout the heliosphere, at least to within $26.9 R_S$. \citet{badman20b} came to a similar conclusion through the analysis of PFSS model solutions for encounters 1-5. As such, the open flux problem must be resolved by processes or uncertainties occurring closer to the Sun. As outlined by \citet{linker17a} and \citet{riley19b}, several possibilities could, in principle, resolve this mismatch, and it is likely that a combination of factors play a role. One of the promising ideas is that an unobserved concentration of magnetic flux lies at the solar poles and provides the ``missing'' open flux to the heliosphere \citep{riley19b}. This may, at least indirectly, be supported by the results here. In particular, the peak values of the polar fields used to drive the WTD model solution for P1 (Figure~\ref{ts_psp_p1} (red)) were only 4.7 G, whereas the peak values in the prediction (Figure~\ref{ts_psp_p1} (magenta)), while more spatially localized, reached up to 65 G. Of course other ideas have also been proposed, including that of \citet{reville20a}, who argued that if the amplitude of the waves used to heat the corona and accelerate the solar wind are added to the radial and total magnetic field kinematically, this would bring the model results into agreement with the measurements. Ultimately, the extent to which polar flux may resolve this ``missing flux’’ problem will be addressed by Solar Orbiter. During its extended mission, it will reach $34^{\circ}$ heliolatitude and such flux, if it exists, should be clearly visible by the Polarimetric and Helioseismic Imager (PHI) onboard.

Finally, we used the model results to explore the global structure of the inner heliosphere during each of the first four perihelia. This allowed us to provide an explanation for the better matches between models and observations for P1 and P3, due to the sharp fold in the HCS and, hence, more separation of the spacecraft from the HCS for large parts of these intervals. On the other hand, during P2 and P4, the spacecraft were often skimming through, or adjacent to the current sheet, making model predictions very sensitive to the precise location of the HCS.  While current solar wind conditions remain extremely quiet, as we follow the ascending phase of the solar activity cycle, such global pictures of the properties and structure in the heliosphere will become ever more useful for interpreting more complex in situ structure. 

This study suggests several possible avenues for future investigations. First, and as noted above, a natural extension to the modeling pipeline would be to assess a fully thermodynamic approach, where all the boundary values from the coronal solution are used to drive the heliospheric model. Based on earlier comparisons, we anticipate that this would, at least initially, yield worse comparisons with in situ measurements; however, this remains to be determined. Second, a systematic sequence of parametric studies, again targeting a smaller number of intervals, such as those studied here, where model parameters and boundary conditions are systematically optimized to produce the best matches with observations. As noted previously (e.g., \citet{riley12e,riley19b}), however, boundary conditions, and, in particular, the radial component of the magnetic field, likely play a -- if not the -- crucial role in the quality of the solution. With the availability of PSP measurements and, most recently, Solar Orbiter data, together with Stereo-A and Wind/ACE, we can now assess the fidelity of the solution at multiple points in longitude, and, eventually latitude, potentially allowing us to estimate what the intrinsic limitations may be in our ability to reproduce the measurements. Finally, and related to this, systematic studies aimed at estimating the contribution of polar fields to in situ measurements, as well as improved models for these values (e.g., ADAPT) may further reduce the uncertainty in model predictions, although, ultimately, the most substantial gains will come from direct measurements of the photospheric field both from polar regions as well as at different heliocentric longitudes, that is, simultaneous $4 \pi$-steradian coverage of the Sun's magnetic field. 

\begin{acknowledgements}
The authors gratefully acknowledge support from NASA (80NSSC18K0100, NNX16AG86G, 80NSSC18K1129, 80NSSC18K0101, 80NSSC20K1285, 80NSSC18K1201, and NNN06AA01C), NOAA (NA18NWS4680081), and the U.S. Air Force (FA9550-15-C-0001).
\end{acknowledgements}

%
%


%
%

\end{document}